\documentclass[traditabstract]{aa} 
\usepackage{txfonts,psfig}

\begin{document}

\title{The amazing diversity in the hot gas content of an X-ray unbiased
massive galaxy clusters sample}
\titlerunning{The amazing diversity of ICM properties} 
\author{S. Andreon\inst{1} \and 
Ana Laura Serra\inst{2} \and A. Moretti\inst{1} \and G.
Trinchieri\inst{1}
}

\authorrunning{Andreon et al.}
\institute{
$^1$INAF--Osservatorio Astronomico di Brera, via Brera 28, 20121, Milano, Italy,
\email{stefano.andreon@brera.inaf.it}\\
$^2$Dip. di Fisica, Universit\`a degli Studi di Milano, via Celoria 16, 20133, Milano, Italy
}
\date{Accepted ... Received ...}
\abstract{
We aim to determine the intrinsic variety, 
at a given mass, of the
properties of the intracluster medium in clusters of galaxies.
This requires a cluster sample  
selected independently of the intracluster
medium content for which reliable 
masses and
subsequent X-ray data can be obtained.
We present one such sample, consisting of  
34 galaxy clusters selected independently of their 
X-ray properties in the nearby ($0.050<z<0.135$) Universe
and mostly with
$14<\log M_{500}/M_\odot \lesssim 14.5$, where masses are
dynamically estimated. 
We collected the available X-ray observations from the archives and then
observed the remaining clusters with the low-background Swift X-ray
telescope, which is extremely useful for sampling a cluster population
expected to have low surface brightness.
We found that clusters display a large range (up to a factor 50) in
X-ray luminosities within $r_{500}$ at a given mass, 
whether or not the central emission ($r<0.15 r_{500}$) is excised,
unveiling a wider cluster population than seen in 
Sunayev-Zeldovich surveys or inferred from the population seen in
X-ray surveys. The measured dispersion is $0.5$ dex in $L_X$ at
a given mass.
}
\keywords{  
Galaxies: clusters: intracluster medium ---
X-ray: galaxies: clusters ---
Galaxies: clusters: general --- 
Methods: statistical --- 
}

\maketitle

\section{Introduction}

The X-ray observations of the hot intracluster medium (ICM) 
of galaxy clusters provide quantities such as its mass,
temperature ($T$), and X-ray luminosity ($L_{X}$).
The analysis of the scaling relations between these physical quantities
gives considerable insight into the physical processes occurring in the ICM (e.g.,
Rosati et al. 2002; Voit 2005). Scaling relations provide the basis for   
cosmological estimates based on 
galaxy clusters (e.g., Vikhlinin et al. 2009). 

Scaling relations have generally been derived for X-ray selected
samples. However, it is becoming 
generally appreciated 
(Pacaud et al. 2007;  Stanek et al. 2006; Pratt et al. 2009;  
Andreon, Trinchieri \& Pizzolato 2011; Andreon \& Moretti 2011; 
Eckert et al. 2011; Planck Collaboration 2011, 2012;
Maughan et al. 2012) that X-ray selected
clusters offer a biased view of the 
cluster population
because, at a given mass, brighter-than-average clusters
are easier to select and include in a sample, 
while fainter-than-average
clusters are easily missed.
In X-ray selected samples, the amplitude of the bias cannot be
determined from the data (and selection function) alone, and
input from non-X ray selected samples is needed to break the
degeneracy between the intercept and the scatter of the
scaling relations.  Alternatively, a strong prior on one of the
parameters, for example based
on numerical simulations, should be adopted (e.g., Vikhlinin et al. 2009). 
In spite of the interest on the subject and 
the general awareness  of the biases induced by the X-ray selection,
a sample of clusters observed in X-rays whose
selection is, at a given mass, independent of the cluster
ICM content is still missing in the literature.  In this
respect, Sunayev-Zeldovich (SZ) based surveys do not seem to represent an alternative, 
given that they also use
the ICM to select the clusters, thus, they may miss clusters with low gas fractions.

Our current (and future) understanding of the complex physics involved in the
cluster formation (see, e.g., Maughan et al. 2012), as well as the cosmological power of
surveys, are severely affected by these systematics that can be understood only with the aid
of cluster samples that are not X-ray selected. Future cosmological surveys (e.g., 
e-Rosita), designed to supply large amounts of data from clusters, will 
benefit from an accurate understanding of the biases
associated with the X-ray selection because the biases  
can only be understood with the aid of 
cluster samples that are {\it not} X-ray selected.

A population of clusters under-represented in 
X-ray selected samples have been already discovered among SZ-selected clusters 
(Planck collaboration 2011, 2012). 
The size of the missing population is not yet quantified; moreover, 
the SZ selection is not a satisfactorily alternative because being based on the 
ICM content it
can miss clusters with low gas fractions.
Andreon \& Moretti (2011) studied a color-selected cluster sample and
found a large variety ($0.5$ dex) of X-ray luminosities
at a given richness. Given that richness correlates
tightly with mass (e.g., Andreon \& Hurn 2010; Andreon 2015), this suggests
that there is a large variety in the $L_X$ at a given mass. So
far, however, this is based on indirect evidence because 
the mass of the studied clusters has not been
directly measured.

To firmly establish the variety of intracluster medium properties
of galaxy clusters of a given mass and to break the degeneracies
between the parameters of the X-ray cluster scaling relations is paramount
to assemble cluster samples having three
features: 1) observed in X-ray; 2) with known mass; and 3) whose
selection is, at a given cluster mass, independent of the intracluster
medium content. If masses are derived from the X-ray data, then 
one should account for the covariance between X-ray luminosity
and mass induced by the double use of the same photons.

The purpose of this work
is to build, observe, and analyze one such cluster sample. 
To this
end, in 2011 our team started an observational program with the X-ray telescope on board of Swift,
dubbed X-ray unbiased cluster survey (XUCS). 
With this program, we were able to uncover a wider range in luminosities relatively 
to what expected on the basis of the Representative XMM-Newton Cluster Structure
survey (REXCESS), or even Plank selected clusters.
The impatient reader may want to
quickly browse \S 2 (sample selection), read \S 3 (which illustrates the main
conclusion of the paper with an example),
skip the technical analysis in \S 4, and read the results based on the whole
sample in \S 5.

\begin{figure}
\centerline{
\psfig{figure=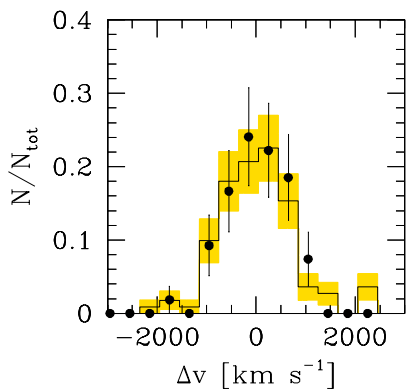,width=6truecm,clip=}
}
\caption[h]{Velocity distribution of the galaxies in clusters CL3013 (histogram)
and CL2007 (points). error-bars to the histogram are
shown with the shading.
In this purely illustrative figure, 
error-bars are simply $\sqrt n$-based. 
}
\end{figure}

Throughout this paper, we assume $\Omega_M=0.3$, $\Omega_\Lambda=0.7$, 
and $H_0=70$ km s$^{-1}$ Mpc$^{-1}$. 
Results of stochastic computations are given
in the form $x\pm y$, where $x$ and $y$ are 
the posterior mean and standard deviation. The latter also
corresponds to 68 \% intervals because we only summarize
posteriors close to Gaussian in this way.

\section{Sample selection}

Our sample, named XUCS, consists of 34 clusters in the very nearby universe 
($0.050<z<0.135$) extracted from the C4 catalog (Miller et al. 2005)
in regions of low Galactic absorption.
The C4 catalog   
identifies clusters as overdensities in the local Universe
in a seven-dimensional space (position, redshift, and
colors; see Miller et al. 2005 for details) using
SDSS data (Abazajian et al. 2004). Among all C4
high-quality clusters ($>30$ spectroscopic members within a radius of 1.5 Mpc, 55 on average) 
with a (dynamical) mass $\log M>14.2$ M$_\odot$ (derived from 
the C4 $\sigma_v>500$ km/s),
we selected the 34 nearest (to minimize exposure times) which are also
smaller than the Swift XRT field of view
($r_{500}\lesssim 9$ arcmin).

The XUCS cluster sample is listed in Table~1.
There is no X-ray
selection in our sample, meaning that 1) the probability of inclusion of the cluster in the
sample is independent of its X-ray luminosity (or count rate), and that 2) no cluster is kept or
removed on the basis of its X-ray properties%
\footnote{Our original
program targeted a slightly larger cluster sample. However, 
because of the approximation in the coordinates listed in the optical C4 catalog, 
the center of a few clusters fell
too close to XRT field-of-view boundaries, or 
in some cases, was missed altogether.
Furthermore, the dynamical analysis of two of the original
clusters
shows that they are two parts of the same dynamical entity
at the $r_{200}$ scale, making the
derived masses of the individual clusters unreliable. 
For this reason, we 
removed these two targets from the sample.}.

As detailed in Table 2, many of our sample clusters were already 
detected by the previous generation of X-ray telescopes and surveys
(e.g., REFLEX and NORAS; Bohringer et al.
2000, 2004), some are listed in the Abell et al. (1989) catalog and some 
are also in the Planck cluster catalog (Planck collaboration 2015).
This information, not used to select the sample, is given for
completeness  and it also illustrates that the clusters selected are
real structures present in other catalogs.

We collected the few X-ray observations present in the XMM-Newton or Chandra archives
and we observed 28 clusters with Swift 
(individual exposure times 
between 9 and 31 ks, total exposure time $\sim 0.35$ Ms, see Table~1). Swift has
the advantage of a low X-ray background (Moretti et al. 2009), 
making it extremely useful for sampling a cluster
population expected to have low surface brightness (Andreon \& Moretti 2011).

\begin{figure}
\centerline{
\psfig{figure=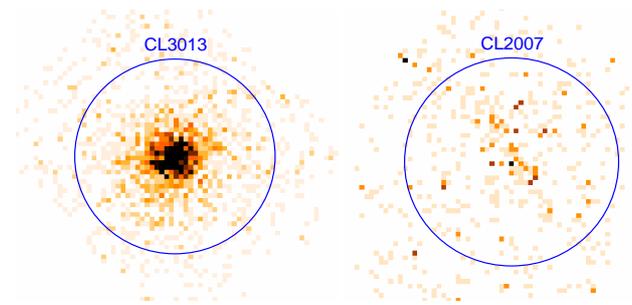,width=9truecm,clip=}
}
\caption[h]{Binned X-ray image of CL3013 and CL2007 in the [0.5-2] keV band
observed for about 9.5 ks each with XRT on Swift.
These clusters have identical masses, velocity dispersions, and redshift,
but X-ray luminosities differing by a factor
15. The circle shows the 0.83 Mpc radius (the $r_{500}$ value).
}
\end{figure}

\begin{figure*}
\centerline{
\psfig{figure=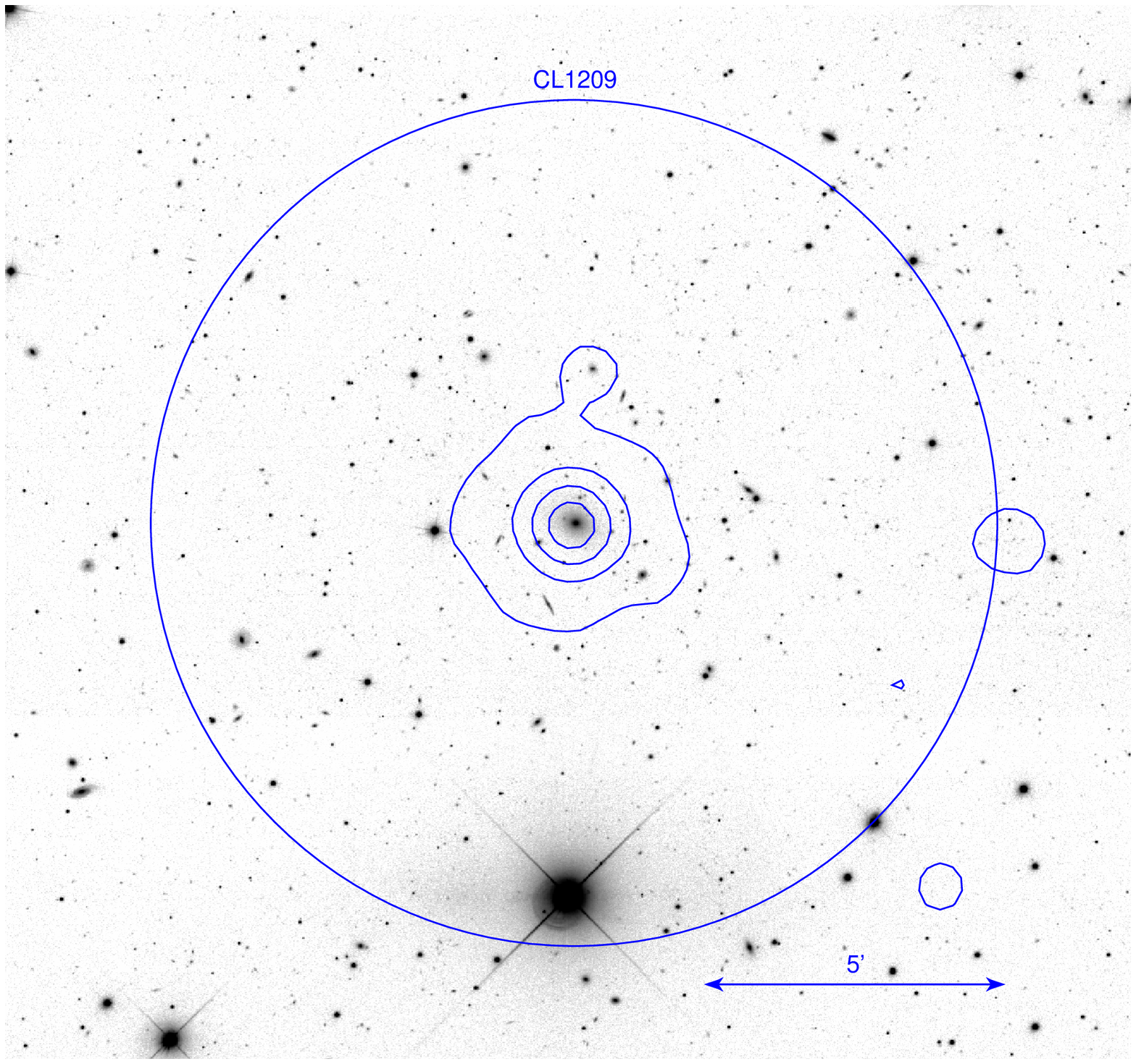,width=8truecm,clip=}
\psfig{figure=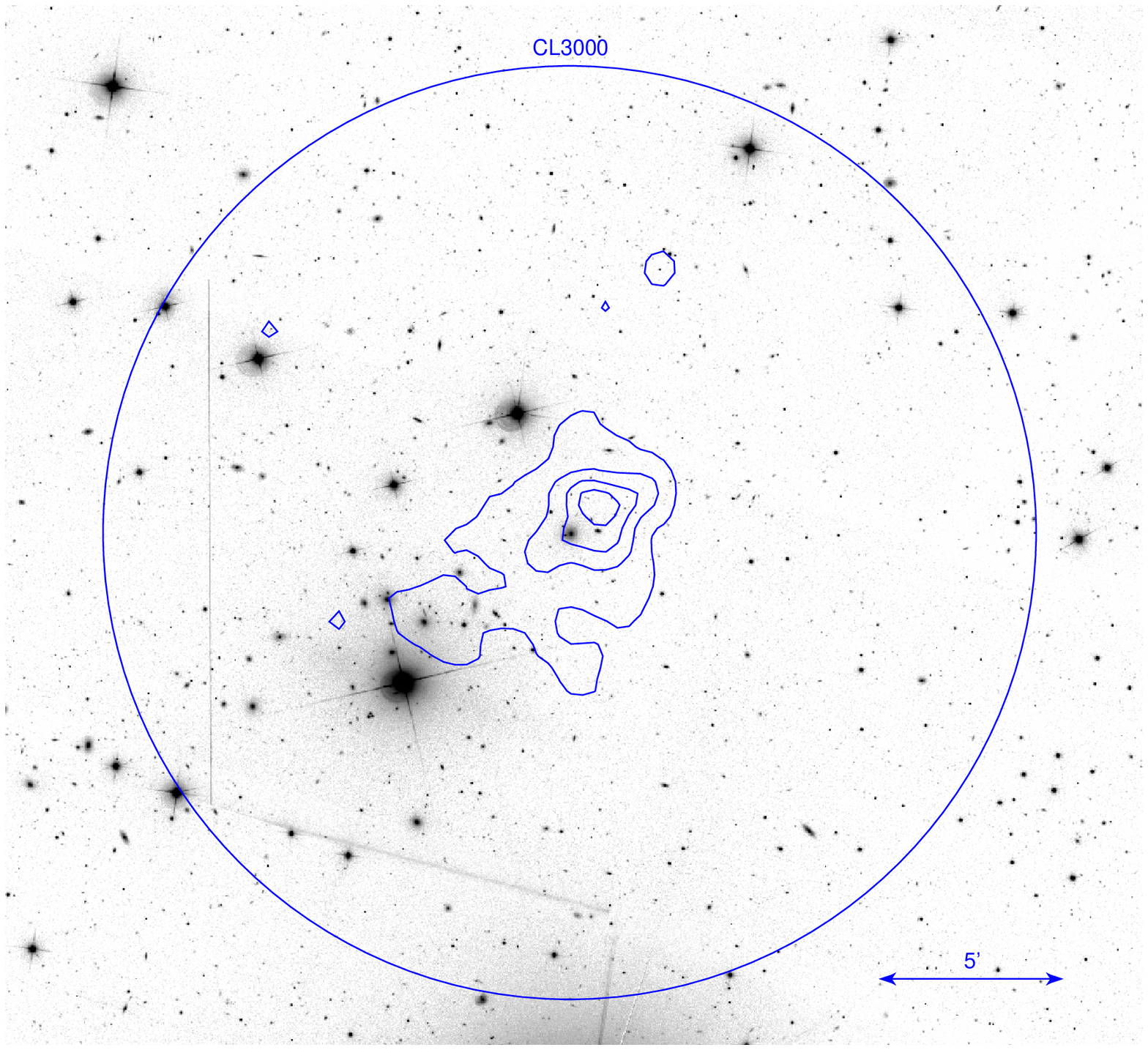,width=8truecm,clip=}
}
\caption[h]{Optical ($r$-band) image of CL1209 (left-hand panel) 
and  CL3000 (right-hand panel) with superimposed X-ray countours
in the [0.5-2] keV band, the latter
smoothed with a Gaussian with $\sigma  \sim45$ arcsec. 
The two clusters are among the X-ray faintest for their mass. 
The circle radius is equal to $r_{500}$. 
North is up, east is to the left.
}
\end{figure*}

\begin{figure*}
\centerline{
\psfig{figure=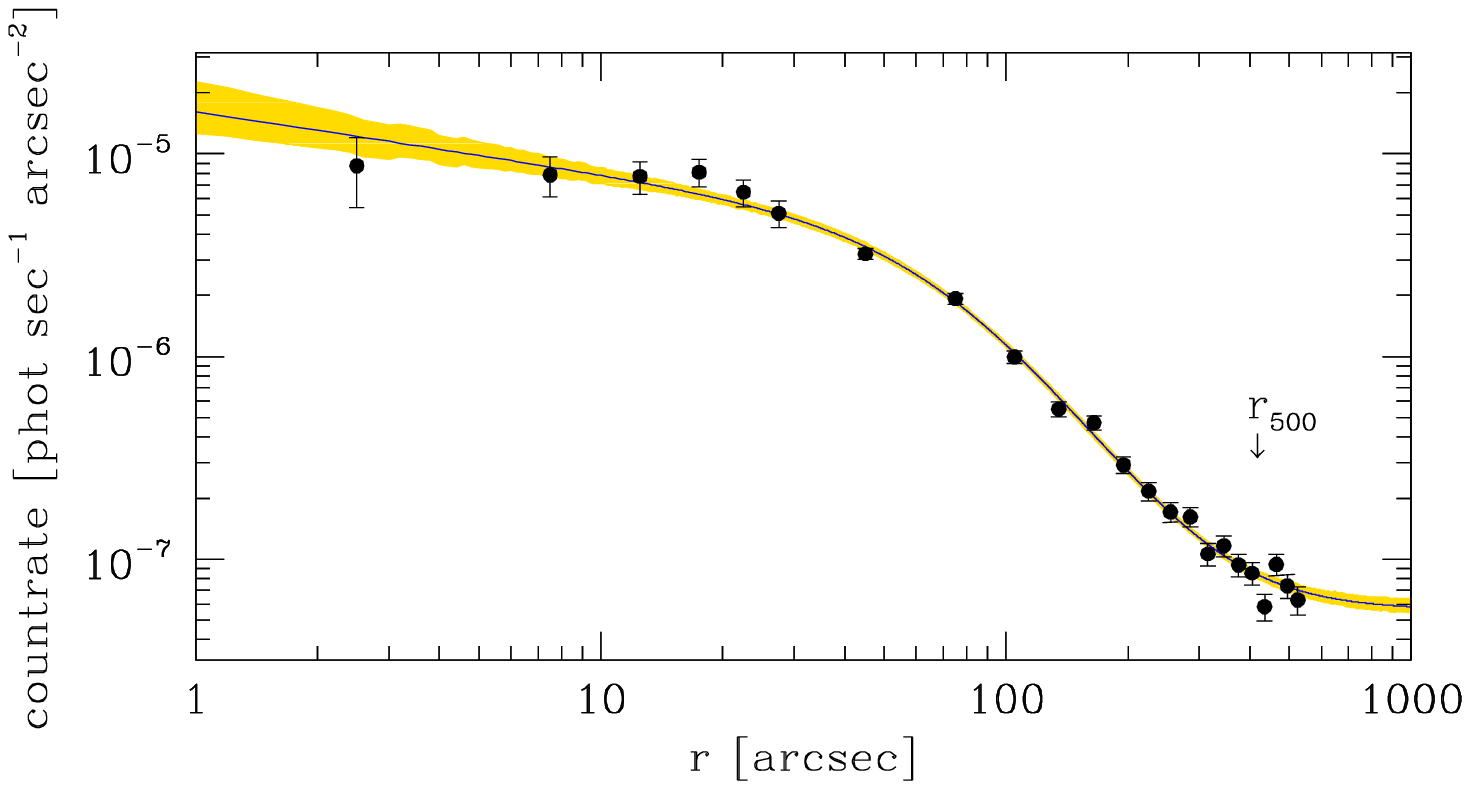,width=9truecm,clip=}
\psfig{figure=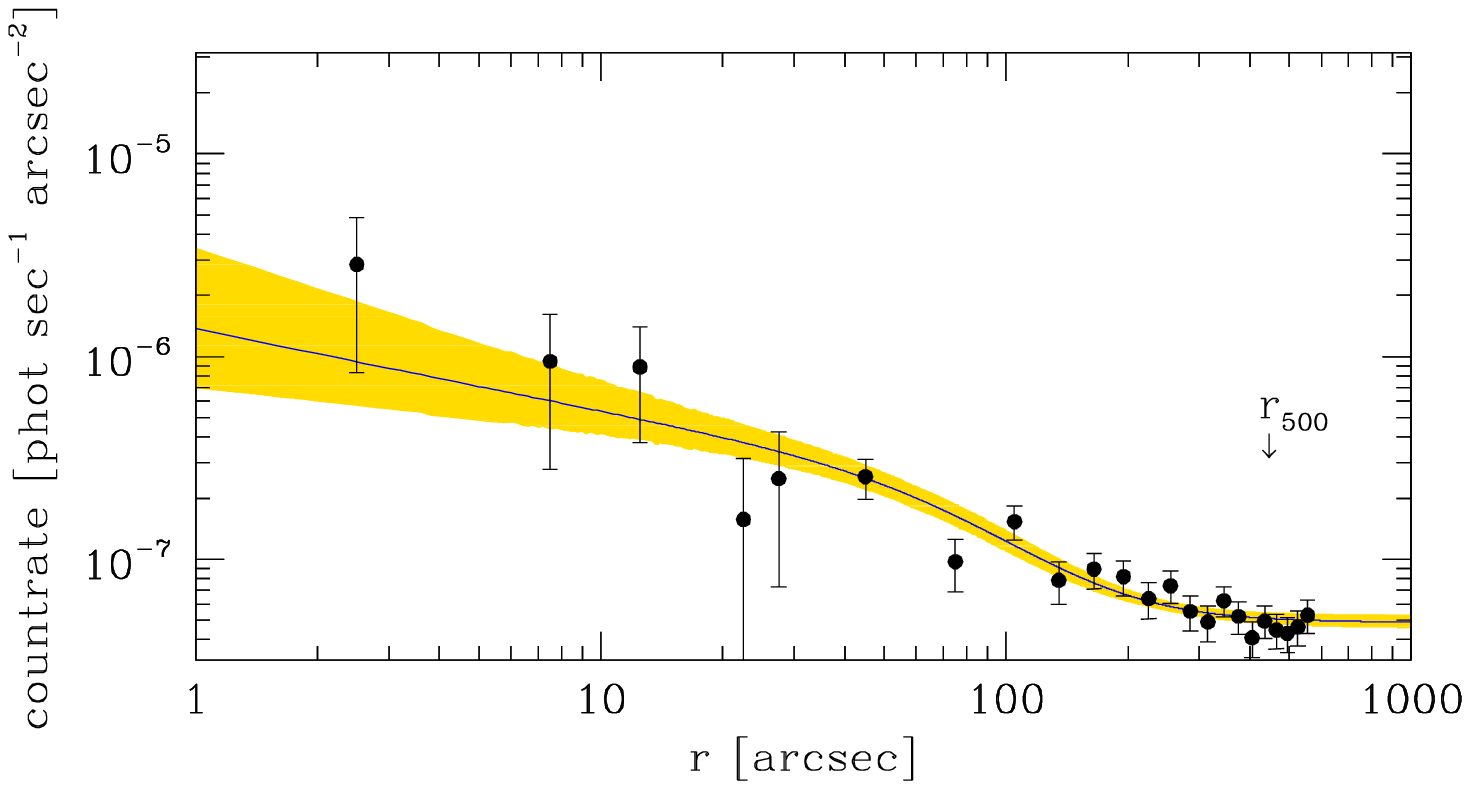,width=9truecm,clip=}
}
\caption[h]{X-ray radial profile ([0.5-2] keV band) of 
the clusters CL3013 (left-hand panel) and CL2007 (right-hand panel). 
The arrow marks the $r_{500}$ value.
}
\end{figure*}

\section{The straightforward comparison} 

To illustrate the diversity in ICM properties of clusters of a
given mass in the simplest possible way, 
we consider two clusters,
CL3013 and CL2007. We chose these two
clusters because 
they have identical masses
($\log M_{500}/M_\odot =14.28$) and velocity dispersions ($600$ km/s; see
Fig.~1);
very similar redshifts ($z=0.115$ and $0.109$); similar number of 
galaxies within the caustics ($\sim 150$); and X-ray data of the same depth ($\sim 9.5$
ks with Swift XRT; see Table 1)  in directions having the same Galactic absorption.
The Swift XRT images shown in Fig.~2 
indicate that the X-ray emission of the two clusters is vastly different. 
We collected
1027 net photons in [0.5-2] keV band for CL3013, and
65 for CL2007
within 0.83 Mpc (corresponding to the $r_{500}$ value).
The more detailed analysis presented in \S 4.1 confirms the large
(a factor $\sim 16$) difference in luminosities (total and core-excised)
and that the difference is not due to a cool core in CL3013 because
the latter
has a $\sim 10$ times larger count rate at all radii.

As discussed in detail in \S 5,
CL3013 and CL2007 differ too much in 
X-ray luminosity 
to obey the low-scatter scaling relation of 
X-ray selected cluster samples, such as the Representative XMM-Newton Cluster Structure
survey (REXCESS) sample (Boheringer et al. 2007),
even after taking the X-ray selection of the latter into account.
This is the main result of this work, which we now illustrate for the full
sample.

\begin{table*}
\caption{Cluster sample and results of the analysis.}
\scriptsize
\begin{tabular}{l r r r r r r r r r l r r r r r}
\hline
Id & ra & dec & $z$ & $\log M_{500}$ & err & $N$ & $\log M_{500}$ & err & $n$ & Telescope & $t_{exp}$ & 
$\log L_X$ & err & $\log L_X$ & err \\ 
 & & & & $[M_\odot]$ & dex & & $[M_\odot]$ & dex & & & ks & erg s$^{-1}$ & dex & erg s$^{-1}$ & dex \\
(1) & (2) & (3) & (4) & (5) & (6) & (7) & (8) & (9) & (10) & (11) & (12) & (13) & (14) & (15) & (16)\\
\hline
CL2081  & 0.167 & 14.550 & 0.093 &   14.39 & 0.09 & 94 &  13.94 & 0.28 & 11 & Swift & 9.5 & 42.44 & 0.11 &  42.09 & 0.19 \\ 
CL2015  & 13.966 & -9.986 & 0.055 &   14.36 & 0.09 & 214 &  14.27 & 0.12 & 61 & Swift & 9.0 & 42.97 & 0.05 &  42.79 & 0.06 \\ 
CL2045  & 22.887 & 0.556 & 0.079 &   13.86 & 0.10 & 66 &  13.95 & 0.16 & 45 & Swift & 10.1 & 43.20 & 0.03 &  43.11 & 0.04 \\ 
CL2010  & 29.071 & 1.051 & 0.080 &   14.34 & 0.09 & 179 &  14.39 & 0.12 & 92 & Swift & 9.4 & 43.25 & 0.04 &  43.10 & 0.05 \\ 
CL2007  & 46.572 & -0.140 & 0.109 &   14.28 & 0.09 & 164 &  14.22 & 0.14 & 57 & Swift & 9.1 & 42.84 & 0.09 &  42.68 & 0.12 \\ 
CL3023  & 122.535 & 35.280 & 0.084 &   14.05 & 0.09 & 173 &  14.31 & 0.17 & 44 & Swift & 9.6 & 43.08 & 0.04 &  42.88 & 0.05 \\ 
CL3030  & 126.371 & 47.130 & 0.127 &   14.52 & 0.15 & 364 &  14.67 & 0.07 & 156 & Swift & 16.5 & 44.05 & 0.01 &  43.93 & 0.02 \\ 
CL3009  & 136.977 & 52.790 & 0.099 &   13.89 & 0.09 & 72 &  13.39 & 0.24 & 17 & Swift & 8.4 & 42.97 & 0.05 &  42.70 & 0.08 \\ 
CL1209  & 149.161 & -0.358 & 0.087 &   14.03 & 0.09 & 115 &  14.29 & 0.15 & 43 & XMM & 49.4 & 42.54 & 0.02 &  42.23 & 0.03 \\ 
CL3053  & 160.254 & 58.290 & 0.073 &   13.52 & 0.27 & 99 &  13.63 & 0.20 & 31 & Swift & 14.4 & 42.61 & 0.08 &  42.48 & 0.07 \\ 
CL3000  & 163.402 & 54.870 & 0.072 &   14.58 & 0.09 & 180 &  14.18 & 0.15 & 59 & Swift & 8.8 & 43.51 & 0.03 &  43.35 & 0.05 \\ 
CL3046  & 164.599 & 56.790 & 0.135 &   14.62 & 0.09 & 172 &  14.63 & 0.15 & 104 & Swift & 9.0 & 44.46 & 0.01 &  44.37 & 0.01 \\ 
  & &  &  &  &  & &  &  & & Chandra & 7.7 & 44.48 & 0.01 &  44.37 & 0.01 \\ 
CL1033  & 167.747 & 1.128 & 0.097 &   13.83 & 0.12 & 77 &  13.79 & 0.21 & 20 & Swift & 9.0 & 43.15 & 0.04 &  43.05 & 0.05 \\ 
CL1073  & 170.726 & 1.114 & 0.074 &   14.20 & 0.09 & 149 &  14.15 & 0.12 & 63 & Chandra & 21.7 & 43.17 & 0.03 &  43.10 & 0.03 \\ 
  &  &  & &   &  &  &  &  &  & Swift & 14.6 & 43.03 & 0.05 &  42.95 & 0.06 \\ 
CL3013  & 173.311 & 66.380 & 0.115 &   14.28 & 0.13 & 148 &  14.29 & 0.12 & 111 & Swift & 9.7 & 44.05 & 0.01 &  43.89 & 0.02 \\ 
CL1014  & 175.299 & 5.735 & 0.098 &   14.40 & 0.09 & 120 &  14.40 & 0.16 & 51 & Swift & 12.0 & 43.47 & 0.03 &  43.31 & 0.04 \\ 
CL1020  & 176.028 & 5.798 & 0.103 &   13.94 & 0.35 & 29 &  14.61 & 0.14 & 43 & Swift & 6.8 & 43.74 & 0.02 &  43.61 & 0.03 \\ 
CL1038  & 179.379 & 5.098 & 0.076 &   14.21 & 0.09 & 95 &  14.28 & 0.13 & 54 & Swift & 8.7 & 43.43 & 0.03 &  43.37 & 0.03 \\ 
CL1015  & 182.570 & 5.386 & 0.077 &   13.85 & 0.19 & 79 &  14.14 & 0.15 & 37 & Swift & 17.3 & 43.65 & 0.01 &  43.55 & 0.01 \\ 
CL1120  & 188.611 & 4.056 & 0.085 &   14.02 & 0.09 & 50 &  14.33 & 0.21 & 25 & Swift & 10.9 & 42.63 & 0.09 &  42.50 & 0.11 \\ 
CL1041  & 194.673 & -1.761 & 0.084 &   14.31 & 0.09 & 370 &  14.58 & 0.11 & 153 & Chandra & 38.2 & 44.39 & 0.01 &  44.17 & 0.01 \\ 
CL1132  & 195.143 & -2.134 & 0.085 &   13.79 & 0.24 & 42 &  13.70 & 0.17 & 27 & Swift & 9.6 & 42.34 & 0.10 &  42.04 & 0.15 \\ 
CL1052  & 195.719 & -2.516 & 0.083 &   14.31 & 0.09 & 262 &  14.57 & 0.10 & 96 & XMM & 49.5 & 43.65 & 0.01 &  43.39 & 0.01 \\ 
CL1009  & 198.057 & -0.974 & 0.085 &   14.00 & 0.26 & 236 &  14.28 & 0.12 & 56 & Chandra & 20.9 & 43.38 & 0.01 &  43.24 & 0.02 \\ 
CL1022  & 199.821 & -0.919 & 0.083 &   13.86 & 0.12 & 87 &  13.56 & 0.16 & 36 & XMM & 31.2 & 42.55 & 0.10 &  42.55 & 0.10 \\ 
CL3049  & 203.264 & 60.120 & 0.072 &   13.63 & 0.10 & 49 &  13.86 & 0.23 & 32 & Swift & 12.3 & 43.09 & 0.03 &  42.97 & 0.03 \\ 
CL1030  & 206.165 & 2.860 & 0.078 &   14.16 & 0.09 & 86 &  14.12 & 0.21 & 41 & Swift & 13.1 & 42.75 & 0.06 &  42.54 & 0.09 \\ 
CL1001  & 208.256 & 5.134 & 0.079 &   14.30 & 0.11 & 171 &  14.57 & 0.09 & 124 & Swift & 31.0 & 43.87 & 0.01 &  43.79 & 0.01 \\ 
CL1067  & 212.022 & 5.418 & 0.088 &   13.92 & 0.38 & 57 &  14.29 & 0.14 & 42 & Swift & 10.9 & 43.33 & 0.04 &  43.25 & 0.04 \\ 
CL1018  & 214.398 & 2.053 & 0.054 &   13.68 & 0.24 & 123 &  14.27 & 0.10 & 79 & Swift & 9.7 & 42.83 & 0.04 &  42.73 & 0.05 \\ 
CL1011  & 227.107 & -0.266 & 0.091 &   14.23 & 0.17 & 48 &  14.46 & 0.14 & 42 & Swift & 13.7 & 42.96 & 0.05 &  42.77 & 0.07 \\ 
CL1039  & 228.809 & 4.386 & 0.098 &   14.32 & 0.16 & 76 &  14.63 & 0.13 & 68 & XMM & 11.0 & 43.52 & 0.02 &  43.42 & 0.02 \\ 
CL1047  & 229.184 & -0.969 & 0.118 &   14.01 & 0.23 & 121 &  13.81 & 0.23 & 28 & Swift & 8.8 & 43.71 & 0.03 &  43.63 & 0.04 \\ 
 &  &  &  &  &  &  &  &  &  & XMM & 39.6 & 43.78 & 0.01 &  43.70 & 0.01 \\ 
CL3020  & 232.311 & 52.860 & 0.073 &   14.12 & 0.14 & 81 &  14.18 & 0.18 & 53 & Swift & 7.3 & 43.30 & 0.04 &  43.20 & 0.04 \\ 
\hline      
\end{tabular} 
\hfill\break
The table lists: cluster id (col 1); coordinates (cols 2 and 3); redshift $z$ (col 4); caustic masses $M_{500}$ (col 5 and 6);
number of galaxies within the caustics (col 7);
velocity-dispersion based masses $M_{500}$ (col 8 and 9); number of galaxies within $r_{200}$ and $|\Delta_v|<3500$ km/s 
(col 10);  X-ray telescope used (col 11); effective exposure time (col 12); X-ray luminosity $L_X(<r_{500})$ in
the [0.5,2] keV band (col 13 and 14); core-excised X-ray luminosity $L_X(0.1r_{500}<r<r_{500})$ in
the [0.5,2] keV band (col 15 and 16).
\end{table*}

\section{Collective analysis of the full sample} 

In the following, we describe the analysis procedures to compute the 
mass and X-ray luminosity in the [0.5-2] keV band of the clusters of our sample.

\subsection{X-ray analysis}

A few clusters of our sample were observed with
XMM-Newton (EPIC-MOS1 and 2; Turner et al. 2001) and Chandra ACIS-I (Garmire et al. 2003;
see Table 1 for details\footnote{XMM obsid 0206430101 (PI C. Miller), 
0201751301 (PI. P. Mazzotta), 0201902101 
(PI H. Boheringer), and 0653810601 (PI Ming Sun);  
Chandra obsid 4990 and 4991 (PI Bower), 5823 (PI Donahue), and 13376 (PI Murray).}), 
which also reports effective exposure times (after flares
cleaning). The remaining clusters were observed  with
the X-ray telescope (XRT) on board the Swift satellite (Gehrels et al.
2004)\footnote{Swift proposals id 8110004, 9120007, and 1013012 (PI S. Andreon),
about one third of them executed
on Italian time.}. 
We reduced the X-ray data using the standard data reduction procedures 
(Moretti et al. 2009; XMMSAS\footnote{http://xmm.esac.esa.int/sas} or
CIAO\footnote{http://cxc.harvard.edu/ciao/}).

As shown in Figs.~2 and 3, clusters in our sample show a variety
of X-ray morphologies. CL3000 shows an offset between  
the galaxy number density 
distribution and the X-ray emission (see Fig.~3), 
a bit like the Bullet cluster (Clowe et al. 2004). 
CL1022 is a bimodal cluster formed by two distinct clumps of 
X-ray emitting gas and galaxies (each one peaked on its own brightest
cluster galaxy; BCG hereafter) of comparable richness and X-ray luminosity. 
The other clusters show a range of morphologies from regular (CL1209, see
Fig.~3) to bimodal/irregulars, as evaluated from the
available number
of photons (the median is $400$ photons within $0.15 r_{500}<r< r_{500}$ 
and $550$ within $r< r_{500}$
in the [0.5-2] 
keV band)\footnote{The radius $r_\Delta$ is the
radius within which the enclosed average mass density is $\Delta$
times the critical density at the cluster redshift. It is computed during the dynamical
analysis in \S 4.2.2.}.

Point sources are detected by 
a wave detection algorithm, and we masked pixels affected by them
when calculating radial profiles and fluxes.
In our analysis, we used exposure maps to calculate the effective
exposure time accounting for dithering, vignetting, CCD defects, gaps, and
excised regions. 

Since Swift observations are taken with different roll angles and sometimes
different pointing centers, the exposure
map may show large differences at very large off-axis angles. To avoid
regions of too low exposure,
we only consider regions where the exposure time is larger than 50\% 
of the central value. Furthermore, we truncate the radial profile when
one-third of
the circumference is outside the 50\% exposure region defined above.
We take the position of the BCG closest to the X-ray peak as cluster center 
(the northern one in case of bimodal cluster CL1022).

\begin{table}
\caption{Literature names}
\scriptsize
\begin{tabular}{l l}
\hline
Id & Literature names \\
\hline
CL2081 &      \\ 
CL2005 & A117, PSZ2G126.72-72.82 \\ 
CL2045 & A208, RXCJ0131.7+0033,  MCXCJ0131.7+0033   \\ 
CL2010 & A279      \\ 
CL2007 & A412      \\ 
CL3023 & A628      \\ 
CL3030 & A655,  RXCJ0825.5+4707,  MCXCJ0825.5+4707,  PSZ2G172.63+35.15 \\ 
CL3009 & XCLASS 1943  \\ 
CL1209 &      \\ 
CL3053 &      \\ 
CL3000 & RXCJ1053.7+5452,   MCXCJ1053.7+5452   \\ 
CL3046 & A1132,    MCXCJ1058.4+5647,  PSZ2G149.22+54.18 \\ 
CL1033 & A1189      \\ 
CL1073 & A1238, RXCJ1122.8+0106,  MCXCJ1122.8+0106   \\ 
CL3013 & A1302, MCXCJ1133.2+6622,  PSZ2G134.70+48.91 \\ 
CL1014 & A1346, PSZ2G261.88+62.85 \\ 
CL1020 &      \\ 
CL1038 & A1424, MCXCJ1157.4+0503   \\ 
CL1015 & ZWCL1207.5+0542, RXCJ1210.3+0523, MCXCJ1210.3+0523   \\ 
CL1120 &      \\ 
CL1041 & A1650, RXCJ1258.6-0145, MCXCJ1258.6-0145, \\
       & Hydra, PSZ2G306.66+61.06 \\ 
CL1132 &      \\ 
CL1052 & A1663, RXCJ1302.8-0230, MCXCJ1302.8-0230, PSZ2G308.64+60.26 \\ 
CL1009 & A1692, RXJ1218.9+0515     \\ 
CL1022 & 2PI0.084J1319.3-0055  \\ 
CL3049 &      \\ 
CL1030 & A1780      \\ 
CL1001 & A1809, MCXCJ1353.0+0509,  PSZ2G339.47+63.56 \\ 
CL1067 & A1864      \\ 
CL1018 & MKW6  \\ 
CL1011 & A2026,  XMMXCSJ1508.4-0015    \\ 
CL1039 & A2048      \\ 
CL1047 & A2051, RXCJ1516.5-0056, MCXCJ1516.5-0056, PSZ2G000.04+45.13 \\ 
CL3020 &      \\ 
\hline      
\end{tabular} 
\end{table}

To measure the surface brightness profile, we extract the
distance from the cluster center $r$ of individual
photon in the [0.5-2] keV band, and we individually fit them (i.e.,
without any radial binning, as detailed in Andreon et al. 2008, 2011; a
similar approach is followed in CIAO-Sherpa), 
with the following generalized $\beta$ model adapted from 
Vikhlinin et al. (2006):
\begin{equation}
S(r) \propto S_0 (r/r_c)^{-\alpha} (1+(r/r_c)^2)^{-3\beta+0.5+\alpha/2} +C \ ,
\end{equation}
where $S_0$, $r_c$, and $C$ are the central surface brightness, core
radius, and the background constant, respectively. The $\alpha$ 
parameter models the power-law-type cusp 
typical of cool-core clusters and is a parameter
requested by the data of some of our clusters. Chandra data
also require us to excise the inner 3-5 arcsec region because of the
BCG emission.
Our fit accounts for vignetting, excised regions,
background level, variation in exposure time, 
and Poisson fluctuations using
the likelihood in Andreon et al. (2008). The parameter
$\beta$ is fixed at $2/3$ because our data cannot constrain
all parameters for all clusters.  For the 
other parameters, we took weak priors, and, in particular, we took a uniform prior
for $S_0$ and $C$, constrained to be positive (to avoid unphysical
values), and a Gaussian prior on $r_c$ with the center equal to $r_{500}/5$
(suggested by Ettori et al. 2015), with a 30\% sigma (i.e., $r_c \in  
(0,r_{500})$ with 99 \% probability).
A posteriori, all (posterior mean) $r_c$ are well within this range.

\begin{figure*}
\centerline{
\psfig{figure=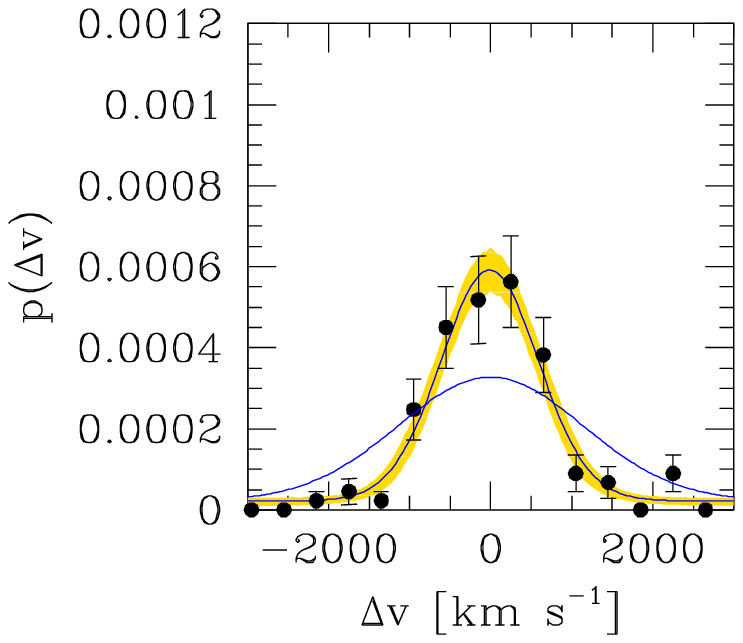,width=6truecm,clip=}
\psfig{figure=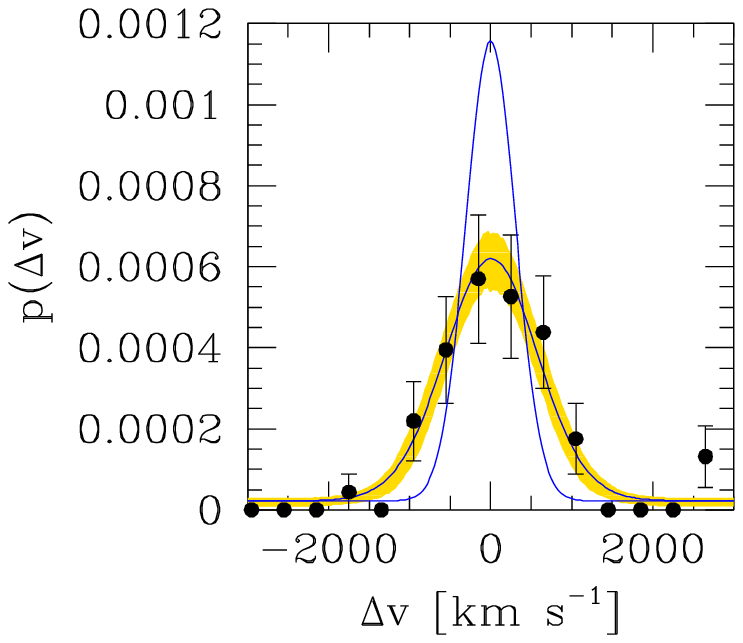,width=6truecm,clip=}
}
\caption[h]{Velocity distribution of the galaxies in the clusters CL3013 (left-hand panel)
and CL2007 (right-hand panel). The solid line
is the mean model fitted to the individual galaxy data in the $r-v$ plane, 
while the shading indicates the 68\% uncertainty
(highest posterior density interval). Points and approximated
error bars are derived by ignoring the radial information, 
binning galaxies in velocity bins, and adopting approximated Poisson   
errors, as is commonly found in the literature. 
The blue curves shows the 
distribution appropriate for their observed X-ray luminosity (see \S 5.4 for details).
}
\end{figure*}

As an example,
Fig.~4 shows the observed surface brightness profile, the fitted
generalized beta model, and its 68 \% uncertainty
for the two 
clusters shown in Fig.~2. CL3013 has no cool core 
and a $\sim 10$ times larger count rate than CL 2007, at all radii,
as mentioned in \S 3.

By integrating the fitted model
within the ranges $0<r<r_{500}$ and $0.15r_{500}<r<r_{500}$ and assuming
a 3.5 keV  temperature, we obtain the X-ray luminosity within
$r_{500}$, $L_{500}=L_X(<r_{500})$, and the core-excised luminosity,
$L_{500,ce}=L_X(0.15<r/r_{500}<1)$, listed in Table~1. 
Adopting $kT=8$ keV would only change the derived luminosities  
by $0.01$ dex. This ensures that our conclusions below are not 
resulting from temperature variations that we have
neglected across the sample.
The smallest $0.15r_{500}$ radius  is 
49 arcsec, much larger than PSF of the used instruments ($\lesssim 8$ arcsec),
so that the contribution of any cool-core flux spilled out 
of the excised region is negligible.

For all but two clusters $r_{500}$ is smaller than the radius
of the (conservatively taken) last point in the surface brightness
profile, $r_{last}$. For CL2015 and CL3000 it is
marginally outside it and we measure the extrapolation 
based on the best-fit model to be
6 to 8 \%.
The error on the uncertain extrapolation is automatically
accounted for by our $L_X$ computation (via Eq.~1).

Although our radial profile model is very flexible in describing the 
surface brightness profile of clusters, it assumes a unimodal
X-ray emission and, therefore, cannot deal with
the bimodal CL1022 cluster. For this cluster,
the X-ray luminosity is derived by simple aperture photometry
centering the aperture on the northern clump (and ignoring  
gaps,
regions unobserved because of other sources, variation in exposure
time, etc.). As a check, 
we derive the X-ray luminosities of CL1132 and CL1015 in a similar way,
i.e., by simply counting photons in the image and with the same
limitations as before.
We found results consistent with those derived by fitting a radial profile, 
which provides more reliable $L_X$ errors however.

Three clusters have data from two X-ray telescopes and we compared the resulting luminosities 
from the different instruments. 
CL3046 has identical luminosities, the other two clusters differ by an
amount that is negligible for our purposes ($\sim0.1$ dex). 
Similarly, the luminosities within the REXCESS (Pratt et al. 2009) $r_{500}$
of the two clusters in common with REXCESS
are consistent.

\subsection{Dynamical analysis}

Data for the dynamical analysis are drawn 
from the 12$^{th}$ SDSS data release (Alam et al. 2015), Galaxy And Mass Assembly (Liskie
et al. 2015; Baldry et al. 2014), 2dF (Colless et al. 2001), 6dF (Jones et al.
2009), Rines et al. (2013),
Hill \& Oegerle (1993), Miller et al. (2002), and Pimbblet et al. (2006).
Both velocity dispersion and caustics analyses below assume the center fixed
at the location of the BCG closest to the X-ray peak.

\subsubsection{Velocity dispersion based masses}

The determination of a dispersion in presence of interlopers/outliers/background
is a well-known problem with a long history: Beers et al. (1990)
reduced the impact of the contaminating population by using lower weights
for objects in the wings of the distribution. 
Andreon et al. (2008), Wojak et al. (2011), and Andreon \& Waever (2015)
modeled the probability that an object belongs to the interesting, or
contaminating, population based on relative velocity from the barycenter, 
which improves the velocity dispersion estimates upon previous 
approaches (Andreon et al. 2008 and Andreon 2010). Here, we
further refine the method with the full cluster centric distance information (i.e., not just
the information that the galaxy is inside the considered radial range) in addition to 
velocity information.

First, we only consider galaxies within $\Delta v<3500$ km/s and $r<r_{200}$
because outside this range galaxies are unlikely to be virialized cluster members.
Second, following previous works, we model the distribution in the velocity space 
as a sum
of a Gaussian (for cluster members) and a uniform distribution (for 
foreground/background galaxies) zeroed outside $\Delta v=3500$ km/s
because we discard galaxies with $\Delta v>3500$ km/s. Third, we model
the distribution in the radial direction as a King profile (for cluster members)
plus a constant (for 
foreground/background galaxies), zeroed at $r>r_{200}$ 
because
galaxies at larger radii are discarded from the sample. 
By fitting a
radial profile, we improve upon previous approaches because we 
use the radial information in full.
The radius $r_{200}$ is, however, unknown.  We
derive it through simple iterations; we first adopt a radius of 20 arcmin, 
and we derive $\sigma_v$ from galaxies within this region, then we convert 
$\sigma_v$ in $M_{200}$ using the $\sigma_v-M_{200}$ relation in Evrard et al. (2008). 
Since $r_{200}\propto M^{1/3}_{200}$, 
we then select only galaxies 
within this $r_{200}$, and repeat the procedure two more times. Convergence is 
already achieved at the first iteration. 

We fit the spatial-velocity model to the (unbinned) position and velocity 
data by means of a Marcov-Chain
Monte-Carlo sampler. We adopt uniform prior
on parameters, zeroed outside the physically acceptable range. Table 1 lists
the derived cluster mass (obtained from the measured $\sigma_v$ using 
the $M_{200}-\sigma_v$ in Evrard et al. 2008, and then in $M_{500}$ assuming
a Navarro, Frenk, and White 1997 profile with concentration $c=5$), 
the number of fitted galaxies, and velocity barycenter. With $c=3$, 
masses would differ by about $0.04$ dex, a difference negligible
compared to our errors.
The median number of fitted galaxies (positions and velocities) 
within $r_{200}$ is
$51$, the interquartile range is $32$. The median mass of the
cluster sample is $\log M_{500} / M_\odot$ is $14.2$, the interquartile range is
0.4 dex.
As mentioned, the probabilistic approach offers more precise
measurements
and more robust estimates of the uncertainties than older approaches
basically because more information in the data is used.

\begin{figure}
\centerline{
\psfig{figure=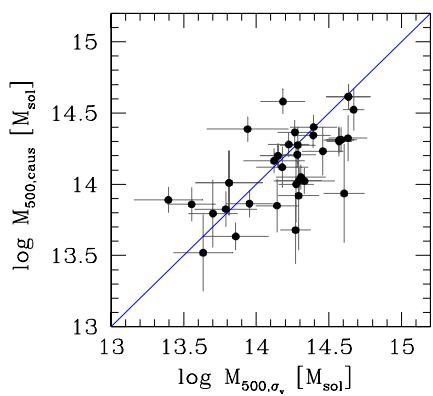,width=6truecm,clip=}
}
\caption[h]{Caustic masses (ordinate) vs velocity dispersion based masses (abscissa).
The blue line indicates equality and it is not a fit to the data.
}
\end{figure}

For a few clusters the assumption of a background population of galaxies
uniform in velocity is not satisfied by the data because of 
the presence of velocity-coherent structures (a filament or a group).
In these cases, we narrowed the radial and/or velocity ranges to
make our assumption valid. 
The caustic analysis described below and an attentive
inspection of the space-phase diagram is quite useful to recognize these
structures.

Fig.~5 shows the observed data, the fitted
model, and its 68 \% uncertainty for the two clusters in Figs~1, 2, and 4.

\subsubsection{Caustics masses}

Caustic masses within $r_{200}$, $M_{200}$, have been derived  following Diaferio \& Geller
(1997),  Diaferio (1999) and Serra et al. (2011). Basically, the caustic technique performs a
measurement of the line-of-sight escape velocity and has the advantage of not assuming virial
equilibrium, assumed instead when estimating velocity-dispersion-based masses. It only uses
redshift and position of the galaxies to identify the caustics on the redshift diagram
(clustrocentric distance vs. line-of-sight velocity), whose amplitude is a measure of the
escape velocity of the cluster. 
The mass
of a spherical shell within the radius $r$ is just the
integral of the square of the caustic amplitude times
the filling factor $F$, taken to be $0.5$ in agreement with
Diaferio \& Geller (1997) and Diaferio (1999).
A byproduct of the caustic
technique is the construction of a binary tree that arranges the galaxies hierarchically,
according to their binding energy. This tool\footnote{CausticApp is publicy
available at 
http://personalpages.to.infn.it/$\sim$serra/causticapp.html.} is 
useful for recognizing possible structures in the considered
field. To force convergence on the cluster under study (instead
of a nearby but distinct cluster in the field), we fixed the cluster
center at the optical/X-ray center.

Table 1 lists the cluster masses $M_{500}$, 
derived from the caustic $M_{200}$ by assuming
a Navarro, Frenk, and White 1997 profile with concentration $c=5$, as for
$\sigma_v$-based masses. 
To account for systematic errors, mostly due to the 
intrinsic triaxial structure of halos,
we adopt as mass error 
the maximum between 20\% and the statistical error (Serra et al. 2011,
Gifford \& Miller 2013). The
median number of members within the caustics is $116$ and the
interquartile range is $45$. The median mass of the
cluster sample is $\log M_{500} / M_\odot$ is $14.2$ and the interquartile range is
0.4 dex.

Eight XUCS clusters have caustic masses also derived in Rines et al. (2006; 
2013). Six out of eight clusters have masses in agreement at 
$1\sigma$ level or better, the remaining two clusters at
$\sim 1.5\sigma$.

While positions and velocities are used by both
the caustic analysis and 
velocity dispersion determinations, 
their use is very different.
For example, the dynamical analysis uses galaxies
within $r_{200}$, while this restriction does not apply
to the caustic technique, which instead mostly uses galaxies at
$r\gtrsim r_{200}$ (and the two $r_{200}$ are not necessarily the
same). 
Furthermore, while we assume a uniform distribution in velocity 
space when we derive masses from the velocity dispersion, the 
caustic technique does not make this
assumption.

Masses inferred from velocity dispersions agree
extremely well with those inferred from caustics (see Fig.~6). The mean $\chi^2$
is $1.3$ and the average scatter is 1.1 times the value expected based
on the quoted errors. Furthermore, we understand the reason for
the difference of the masses of the two clusters with the largest
differences ($\sim 2\sigma$):
CL3000 is likely to have underestimated caustic errors and
the velocity dispersion of CL1018 is likely
overestimated because of a possible background 
velocity structure not accounted for.

\begin{figure}
\centerline{\psfig{figure=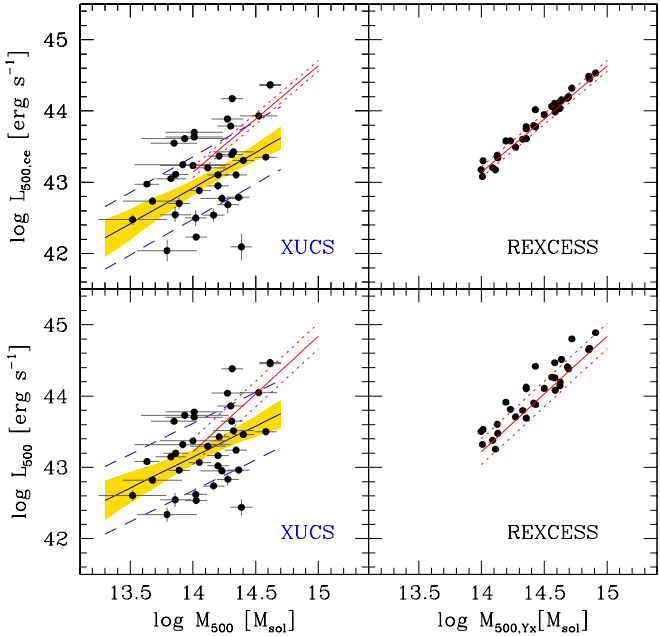,width=9truecm,clip=}}
\caption[h]{X-ray luminosity ([0.5-2] keV band) vs Mass. {\it Top panels:}
core-excised. {\it Bottom panels:} non-core excised. {\it Left-hand panels:} our (XUCS)
sample. {\it Right-hand panels:} REXCESS sample. The solid (red) and dotted lines are
the REXCESS Malmquist-bias corrected mean and estimated
scatter. The solid blue line surronded by a yellow shading indicates our fit to our data.
The shading indicates 
its 68\% uncertainty and the dashed blue dashed lines indicate the 
fit model $\pm$ the intrinsic
scatter.
}
\end{figure}

\section{X-ray luminosity vs masses}

The left-hand panels of Fig.~7 show the X-ray luminosity $L_{500}$ (bottom 
panels) and the core-excised X-ray luminosity $L_{500,ce}$ 
(top panels) vs mass for our
cluster sample. 

We fit the data above with a linear relation with intrinsic scatter
and errors\footnote{The
fitting code is distributed in Chapter 8.4 of Andreon \& Waiver 2015, 
in Andreon \& Hurn 2013 and also
available at
http://www.brera.mi.astro.it/$\sim$andreon/
BayesianMethodsForThePhysicalSciences/model8.4.bug} 
with the JAGS Marcov-Chain Monte-Carlo sampler, which adopts
weak priors on parameters. For core-excised luminosities
we found
\begin{equation}
\log L_{500,ce} = (0.82\pm0.33)(\log M_{500}/M_\odot-14.2) + 43.2\pm0.1 \ ,
\end{equation}
with an intrinsic scatter of $0.50\pm0.07$, and a very similar result for
total luminosities
\begin{equation}
\log L_{500} = (0.85\pm0.31)(\log M_{500}/M_\odot-14.2) + 43.3\pm0.1 \ ,
\end{equation}
with an intrinsic scatter of $0.47\pm0.07$. 
Luminosities are expressed in erg s$^{-1}$  and the scatter accounts
for both the noisiness of mass and X-ray luminosity.
These results are robust
to the modeling of the scatter: we find indistinguishable results
by replacing the 
assumed Gaussian scatter with a Student's t distribution with
10 degree of freedom, which is more robust to the presence
of outliers because has more extended tails (Andreon 2012b details how to implement it).
We find the same
scatter when we restrict our attention to $\log M_{500}/M_\odot>14$. 

The right-hand panels of Fig.~7 show X-ray luminosity vs mass, but 
for the REXCESS (Pratt et al. 2009) cluster sample.
This sample is composed of X-ray selected clusters in the range $0.06<z<0.17$.
By construction, plotted
points tend to be above the fit line because the line 
represent the scaling relation inclusive of a correction for
the X-ray selection function. 
The XUCS sample, is, as mentioned, 
X-ray unbiased and largely overlaps REXCESS in redshift
($0.06<z<0.17$ vs $0.05<z<0.14$). The mass ranges are also largely
overlapping: 70\% of XUCS and 50\% of REXCESS clusters are in 
the $14<\log M_{500}/M_\odot<14.5$ range. While the REXCESS sample
tightly obeys the luminosity-mass scaling, the XUCS cluster sample shows
a much larger diversity in luminosity at a given mass. 

The scatter observed in XUCS is 2.7 ($L_{500}$) and 7 ($L_{500,ce}$)
times the one inferred in REXCESS
accounting for the X-ray selection. Therefore,
the observed scatter found in our sample 
cannot be simply explained within the scatter 
derived accounting for the X-ray selection bias. 
Rather, with this sample we have unveiled a cluster 
population exhibiting a larger range of properties for the same mass. 
This is true also if we consider the $\log M_{500}/M_\odot > 14$ mass range 
(see Fig.~7).

The variety seen in XUCS is also larger than that seen 
by Hicks et al. (2013):
the few clusters studied by these authors occupy the low $L_X$ side 
of the REXCESS distribution, while our clusters go well below.
The large variety is indirectly confirmed by Andreon \& Moretti (2011),
who observed a $0.5$ dex scatter at a given richness, which is known to be
a low scatter proxy of mass (e.g., Andreon \& Hurn 2010; Andreon 2015).
That sample, however, lacks direct measurements of mass.

\begin{figure}
\centerline{\psfig{figure=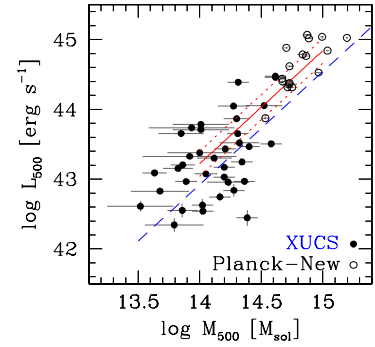,width=5truecm,clip=}}
\caption[h]{X-ray luminosity ([0.5-2] keV band) vs 
mass for our sample (left-hand panel) and the new Planck
clusters (right-hand panel). Masses for Planck clusters are $Y_X$-based,
whereas they are caustics based for XUCS. The solid (red) and dotted lines are
the REXCESS Malmquist-bias corrected mean and mean $\pm$ the estimated
scatter. The long dashed (blue) line is the mean REXCESS scaling 
offset by
0.3 dex describing the lower envelope of Planck clusters. 
Several XUCS clusters are below this envelope. 
}
\end{figure}

\subsection{Is variety correlated to X-ray morphology?}

Inspection of the X-ray images of clusters with low X-ray luminosity
for their mass  do not show any obvious relation between morphology of the X-ray
emission and the location of the cluster in the $L_X-M_{500}$ plane.
For example, CL3000 (see Fig.~3, $0.8$ dex below the mean REXCESS relation) 
is an irregular cluster,
and, based on the displacement between X-ray emission and galaxies, an interacting
cluster. However, CL1001 shows a similar morphology but it falls close to the REXCESS 
mean. Furthermore, while CL2007 (see
Fig.~2, 1.0 dex below the mean REXCESS relation) 
is irregular and of low luminosity for its mass, CL1209 (1.0 dex below the mean REXCESS 
relation) is also
of low luminosity but its X-ray appearance is fairly regular (see Fig.~3). Therefore,
this suggests no obvious relation between the X-ray morphology and position
in the $L_X$-mass plane.  Paper II will more accurately explore this evidence.

\subsection{A variety larger than pointed out by Planck?}

Figure~8 shows the X-ray luminosity vs mass scaling relation for XUCS and for
the subsample of new (i.e., not previously known) Planck clusters (Planck
collaboration 2011). Masses for Planck cluster are
(as REXCESS) based on the pseudo-pressure parameter $Y_X$ 
(Kravtsov et al. 2006). The relation defined by the
Planck clusters also shows
a larger variety than REXCESS (Planck collaboration 2011). While
they do not reach the amplitude displayed by this sample,
the more extreme point is just $0.5$ dex ($<3\sigma$) below the best-fit
REXCESS relation.
In XUCS, CL2081
($\log M_{500}/M_\odot = 14.4$) is off by 1.4 dex, and almost half of the whole sample 
is more than 0.3 dex fainter than the REXCESS mean
(long dashed line).  Although the two data sets are sampling complementary parts of
the mass function (Planck above, and XUCS below, $\log M_{500}/M_\odot = 14.5$), 
the variety in the X-ray luminosity at a given
mass that we see is larger than observed by Planck, suggesting either that Planck is
only unveiling part of the population unveiled by XUCS or
that the scatter is mass-dependent.
An SZ-selected sample is expected to show a reduced
variety of X-ray properties from numerical
simulations (Angulo et al. 2012) because the ICM that is emitting in X-ray is also
responsible for the inverse Compton effect; this makes the two
observables, and selections based on them, correlated.

\begin{figure}
\centerline{\psfig{figure=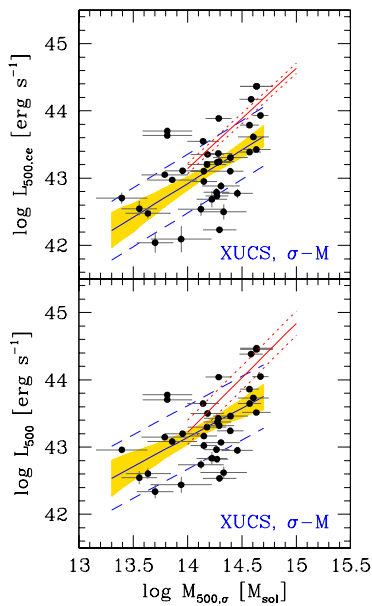,width=5truecm,clip=}}
\caption[h]{X-ray luminosity ([0.5-2] keV band) vs 
velocity-dispersion-based masses. {\it Top panel:} core-excised
luminosities. {\it Bottom panel:} non-core-excised luminosities.
Shading, solid line and
dashed corridor are uncertainty on the mean model, mean model,
and mean model $\pm$ the intrinsic scatter, respectively.
}
\end{figure}

\subsection{Is variety due to wrong X-ray fluxes?}

In this section and in the following sections we investigate possible reasons
for the different
behaviour of the XUCS and REXCESS samples, 
starting with the exploration of possible inaccuracies in the $L_X$ 
determination. 

In Sect.~3 we show two clusters with similar mass but very different
X-ray luminosities. It is very hard to 
explain a difference, in excess of  900 counts, needed to reconcile 
the two luminosities, from a faulty data analysis, in particular, 
when the total number of photons collected in the image of CL2007 
is of about 700, inclusive of background and other sources. 
In Sect.~4.1 we showed that $L_X$ derived by 
aperture counting is consistent with our more elaborate analysis
accounting for excised regions, gaps, variation in
exposure time, etc, and that we obtain consistent values
using different telescopes. Our $L_X$
is also consistent with that derived by REXCESS
for the two clusters in common. Finally, a
lack of knowledge of the X-ray temperature affects
$L_X$ to a negligible level ($0.01$ dex) that is
50 times too small to explain the
observed $0.5$ dex scatter.

To have overly faint clusters we
must have overestimated their backgrounds. A plot of
the $L_X$ residuals vs background value shows that this is not
the case: overly faint clusters do not have larger values of 
estimated background, and, furthermore, 
all XRT background estimates are within 15\%, 
thanks to the stability of the low Swift X-ray background.
To further check the background subtraction, we recompute the
CL1067 X-ray luminosity using the background from
a pointing 31 arcmin away from the cluster. The X-ray luminosity
increases by $0.05$ dex, 1.5 times the quoted
error and ten times smaller than needed to explain the observed $L_X$ scatter at 
a given $M$. To summarize, we are confident that 
the cluster X-ray luminosities are precisely measured and cannot be 
the source of the observed scatter.

\subsection{Is variety due to wrong cluster masses?}

The $L_X$-mass scaling explored thus far uses masses derived from
caustics, which does not assume
the hydrostatic equilibrium 
nor a relaxed status for the cluster. 
The noise and bias of caustic masses have been characterized by a few
independent teams who use different simulation methods and assumptions. Serra et
al. (2011) used hydrodynamic simulations and find a bias of $<$5\% at about $r_{200}$
($\sim$ 10\% at most radii), uncertainties on individual determinations of about
25\% (at about $r_{200}$), and errors in agreement with the uncertainties above.
Gifford \& Miller (2013), via N-body simulations, also found a small bias
(smaller than 4\%) and uncertainties on individual determination of about
10-15\%. Gifford, Miller, \& Kern (2013) confirmed and extended these results with
N-body simulations populated by galaxies using different semianalytic
prescriptions. Svensmark et al. (2015), using N-body simulations, found no bias and
a $\sim 15$\% scatter. To explain the observed variance in $L_X$ at a given
mass, caustic masses should have a scatter of 300\% (in addition to the quoted errors).

Observations also indicate that caustic masses cannot be much noisier than their quoted error
(i.e., cannot have have understimated errors) because they correlate with almost no
scatter with richness (Andreon 2015), which in turns correlates with almost no scatter with
weak-lensing mass (Andreon \& Congdon 2014).
This latter correlation adopts fixed-aperture masses and, therefore, the source of the tight relation
cannot be an effect introduced using mass-related apertures such as $r_{200}$.

In spite of the evidence provided by previous authors that caustic masses and error are accurate
and well understood, and in addition to our accounting of the mass noisiness in our analysis,
we want to test our results further by considering masses derived from the galaxy
velocity dispersion.

Figure~9
shows the luminosity-mass scaling relation using these masses.
By fitting the
data we found scaling relation consistent with those measured with caustic
masses, namely:
for core-excesed luminosities
\begin{equation}
\log L_{500,ce} = (1.00\pm0.28)(\log M_{500}/M_\odot-14.2) + 43.1\pm0.1 \ ,
\end{equation}
with an intrinsic scatter of $0.44\pm0.07$, and for
total luminosities
\begin{equation}
\log L_{500} = (0.85\pm0.31)(\log M_{500}/M_\odot-14.2) + 43.3\pm0.1 \ ,
\end{equation}
with an intrinsic scatter of $0.47\pm0.07$. In particular,
both scatter estimates are in agreement with those derived with
caustic masses, confirming the variety found there.

These X-ray luminosities are still measured within the $r_{500}$
radius derived from caustics. In order to check that the scatter
is not due to mismatched apertures, we also adopted 
radii derived from velocity dispersions (see \S~4.2.1). The
large variety of $L_X$ values at a given mass is still there.
Finally, we checked that using the Munari et al. (2013)
relation in place of the Evrard et al. (2008) relation
(to convert velocity dispersions in masses)
does not affect the scatter at all. Indeed,
it can be easily shown mathematically that the scatter is unaffected
by whatever  velocity dispersions vs masses relation is used, provided
it is a continous function.

To quantify the amplitute of the error on $\sigma_v$ needed to explain the
scatter of the $L_X$-mass relation, let us consider again the clusters CL3013 and 
CL2007.
The slope of the scale relation between core-excised luminosity and mass is $1.49$ 
(Pratt et al. 2009), the slope between mass and velocity dispersion 
is close to $3$ (e.g., Evrard et al. 2008; Munari et al. 2013). Therefore,
from scaling relations, the velocity dispersion of CL3013 should be
almost twice the velocity dispersion of CL2007 (as shown by the blue solid line in Fig.~5,
the velocity distribution that is expected to match the cluster X-ray luminosity). Clearly,
this is ruled out by the data.

\subsection{Could be the REXCESS scatter underestimated?}

As already remarked, several works (Plack Collaboration 2011; Andreon \& Moretti 2011; Hicks et al.
2013) find a more variegate cluster population than inferred from the REXCESS sample, i.e., that the
REXCESS scatter underestimates the total scatter in $L_X$ at a given mass. Here we want to focus on a
relevant assumption at the base of the mass derivation.
Pratt et al. (2009) derive masses  from $Y_X=M_{gas} T$, where
the product is largely driven by the change in $M_{gas}$ because $T$ changes by about 
25\% for most of the
REXCESS sample, while $M_{gas}$ variations are of about 100\%. The gas mass
and $L_X$ are derived from the same
X-ray photons, in particular, from two closely related
radial profiles (of $n_e$ and $n^2_e$), 
inducing a covariance between the $Y_X$-derived
mass and $L_X$ 
because of the double use of the same
information. In fact,
Maughan et al. (2007) find a linear relation
between $\log M_{gas}$ and $\log L_X$ with a
negligible (3 \%) scatter. Said simply, 
a cluster brighter/fainter for its mass also has a
larger/smaller $Y_X$--based mass, moving mostly along
the $L_X-M_{Y_X}$ relation, giving a false impression of a small
variance in the REXCESS sample. Since only
the orthogonal component of the scatter has been measured and
reported by Pratt et al. (2009), the orthogonal scatter
underestimated the $L_X$ scatter at a given mass.

The above (neglected) covariance is a general
feature of the $L_X$ vs $M_{Y_X}$ analysis, and applies to
$Y_X$ based Planck masses in Fig.~6: the variety in
the Planck sample may be larger than it appears in Fig.~6.
Instead, caustic masses are not derived using X-ray photons. This kind
of covariance is absent for our mass-$L_X$ determination and
our scatter is not underestimated for that reason.

\section{Summary and Discussion}

To firmly establish the variety of intracluster medium properties
of clusters of a given mass and to break the degeneracies
between the parameters of the X-ray cluster scaling relations is important
to assemble cluster samples with the following three
features: 1) observed in X-ray; 2) with known mass; and, 3) whose
selection is, at a given cluster mass, independent of the intracluster
medium content. Having masses derived independently of the X-ray data
is also useful to neglect the covariance induced by the double use
of the X-ray photons. The sample, XUCS, presented in this paper
offers, for perhaps the first time, an X-ray unbiased view of the cluster
population because the cluster sample is assembled independently
of the ICM content; at a given mass, clusters are in our sample independent
of their X-ray luminosity. It consists of
34 massive clusters ($\log M_{500}/M_\odot > 13.5$ with
70\% of the clusters within $14<\log M_{500}/M_\odot \lesssim 14.5$)
in the nearby Universe
($0.05<z<0.135$), observed in X-ray, with abundant spectroscopic data
($116$ galaxies within the caustics on average),
and reliable masses.

XUCS reveals an unexpected variety in X-ray luminosity at 
a given mass, up to a factor 50. The scatter in X-ray
luminosity (total or core-excised) is $0.5$ dex.
The observed range (and spread) in XUCS is about
five times larger than the spread inferred from REXCESS 
after accounting for the X-ray
selection of the latter. We are not just recovering those
clusters known to be missed in X-ray selected samples,
but a broader population missed altogether by REXCESS.
The variety we observe in XUCS also seems larger than in the 
new Planck cluster sample. 
This suggests that either Plank is only partially probing the full population 
seen by XUCS, although the two surveys cover complementary parts 
of the cluster mass function, or that the scatter in $L_X$ at a given mass 
is mass dependent (larger at lower masses).

Various tests have shown that our X-ray luminosities are reliable and not the
cause of the observed variety. Furthermore, by using core-excised 
X-ray luminosities, we verified that the variety is not due to a central
luminosity excess. Masses are derived independently of the X-ray
data. In particular, we use caustic masses,
which have the advantage of not assuming virial equilibrium nor a
relaxed status for the cluster, or
velocity dispersion-based masses, which have reliable
errors. In both cases, we found the same 
variety of ICM properties at a given mass.

We known that clusters show a variety in their ICM properties, and in fact
there are many individual examples of the effects of AGN
feedback (e.g., Voig \& Donanhue 2005), star formation (e.g., Nagai 2006),
mergers (Rowley et al. 2004), radiative cooling 
and other nongravitational effects on the ICM properties. 
However, the frequency with which these processes occurs 
in the general cluster population cannot be inferred
from an etherogenous collection of clusters and 
requires an X-ray 
unbiased sample. In XUCS, we found a scatter of
$0.50$ dex at a given mass, indicating a much more etherogenous population
than previously believed based on X-ray or SZ selected
samples. It is left to a future paper to
study the implications of the large scatter above on our understanding
of the complex ICM physics. Here we just
mention that past numerical simulations
attempted to infer which physical
processes are running in clusters (and
the sub-grid physics) by matching slope, intercept, and
scatter of observed and simulated
scaling relations (e.g., Young et al. 2011). 
Our larger observed variety prompts some
stochasticity (across clusters) of the importance of the different
processes, or asks us to enlarge the range of values of parameters
addressing the subgrid physics currently kept fixed or bounded
in a narrow range (e.g., Pike et al. 2014).

It is believed that X-ray luminosity is a reliable mass proxy with
controllable scatter (e.g., Maughan 2007, Pratt et al. 2009), which has important
implications for cluster surveys such as Planck and eROSITA, and ultimately
for the use of the cluster population for cosmological purposes. 
At first sight, the
variety seen in our sample severely reduces the use of
X-ray luminosity, including the core-excised one,
no longer a mass proxy of moderate scatter. This has an
obvious consequence on inferring mass from the observable, and a subdle effect
on the knowledge of the selection function, which is central for cosmological
tests, of X-ray selected samples. Since the scatter is large,
it is hard to characterize the fraction of the population seen by
the survey. On the other end, it might be possible to identify
subsamples of clusters with a lower variance at a given mass,
for example, those with $\log M_{500}/M_\odot > 14.5$, which may show a
smaller spread, as Fig.~8 may suggest, and perform
cosmological tests using these. We defer a more detailed
discussion to a later paper, when the results of our current X-ray
observing campaign on more massive clusters will be completed.

\begin{acknowledgements}
ALS acknowledges useful discussions with Antonaldo Diaferio.
For the standard SDSS III and GAMA
acknowledgements see: 
https://www.sdss3.org/collaboration/boiler-plate.php and
http://www.gama-survey.org/pubs/ack.php
\end{acknowledgements}

{}


\begin{thebibliography}{}


\bibitem[Abazajian et al.(2004)]{2004AJ....128..502A} 
Abazajian, K., Adelman-McCarthy, J.~K., Ag{\"u}eros, M.~A., et al.\ 2004, \aj, 128, 502 

\bibitem[Abell(1958)]{1958ApJS....3..211A} 
Abell, G.~O.\ 1958, \apjs, 3, 211 

\bibitem[Alam et al.(2015)]{2015arXiv150100963A} 
Alam, S., Albareti, F.~D., Allende Prieto, C., et al.\ 2015, ApJ, submitted 
(arXiv:1501.00963) 


\bibitem[Andreon(2010)]{2010bmic.book..265A} 
Andreon, S.\ 2010, in Bayesian Methods in Cosmology, 
	eds by M. Hobson, A. Jaffe, A. Liddle, P. Mukeherjee and D. Parkinson.,
        Cambridge University Press, New York, Cambridge, UK

\bibitem[Andreon(2012)]{2012A&A...548A..83A} 
Andreon, S.\ 2012a, \aap, 548, A83 

\bibitem[Andreon(2012)]{2012A&A...546A...6A} 
Andreon, S.\ 2012b, \aap, 546, A6 

\bibitem[Andreon(2015)]{2015A&A...582A.100A} 
Andreon, S.\ 2015, \aap, 582, A100 

\bibitem[Andreon \& Congdon(2014)]{2014A&A...568A..23A} 
Andreon, S., \& Congdon, P.\ 2014, \aap, 568, A23 

\bibitem[Andreon \& Hurn(2010)]{2010MNRAS.404.1922A} 
Andreon, S., \& Hurn, M.~A.\ 2010, \mnras, 404, 1922 

\bibitem[Andreon \& Moretti(2011)]{2011A&A...536A..37A} 
Andreon, S., \& Moretti, A.\ 2011, \aap, 536, A37 

\bibitem[Andreon \& Weaver(2015)]{Springerbook} 
Andreon, S., Weaver, B. \ 2015, Bayesian Methods for the Physical Sciences.
Learning from Examples in Astronomy and Physics, Springer 

\bibitem[Andreon et al.(2011)]{2011MNRAS.412.2391A} 
Andreon, S., Trinchieri, G., \& Pizzolato, F.\ 2011, \mnras, 412, 2391 

\bibitem[Andreon et al.(2008)]{2008MNRAS.383..102A} 
Andreon, S., de Propris, R., Puddu, E., Giordano, L., \& Quintana, H.\ 2008, \mnras, 383, 102 

\bibitem[Angulo et al.(2012)]{2012MNRAS.426.2046A} 
Angulo, R.~E., Springel, V., White, S.~D.~M., et al.\ 2012, \mnras, 426, 2046 

\bibitem[Baldry et al.(2014)]{2014MNRAS.441.2440B} 
Baldry, I.~K., Alpaslan, M., Bauer, A.~E., et al.\ 2014, \mnras, 441, 2440 

\bibitem[Beers et al.(1990)]{1990AJ....100...32B} 
Beers, T.~C., Flynn, K., \& Gebhardt, K.\ 1990, \aj, 100, 32 

\bibitem[B{\"o}hringer et al.(2004)]{2004A&A...425..367B} 
B{\"o}hringer, H., Schuecker, P., Guzzo, L., et al.\ 2004, \aap, 425, 367 

\bibitem[B{\"o}hringer et al.(2007)]{2007A&A...469..363B} 
B{\"o}hringer, H., Schuecker, P., Pratt, G.~W., et al.\ 2007, \aap, 469, 363 

\bibitem[B{\"o}hringer et al.(2000)]{2000ApJS..129..435B} 
B{\"o}hringer, H., Voges, W., Huchra, J.~P., et al.\ 2000, \apjs, 129, 435 

\bibitem[Clowe et al.(2004)]{2004ApJ...604..596C} 
Clowe, D., Gonzalez, A., \& Markevitch, M.\ 2004, \apj, 604, 596 

\bibitem[Colless et al.(2001)]{2001MNRAS.328.1039C} 
Colless, M., Dalton, G., Maddox, S., et al.\ 2001, \mnras, 328, 1039 

\bibitem[Diaferio(1999)]{1999MNRAS.309..610D} 
Diaferio, A.\ 1999, \mnras, 309, 610 

\bibitem[Diaferio \& Geller(1997)]{1997ApJ...481..633D} 
Diaferio, A., \& Geller, M.~J.\ 1997, \apj, 481, 633 

\bibitem[Eckert et al.(2011)]{2011A&A...526A..79E} 
Eckert, D., Molendi, S., \& Paltani, S.\ 2011, \aap, 526, AA79 

\bibitem[Ettori(2015)]{2015MNRAS.446.2629E} 
Ettori, S.\ 2015, \mnras, 446, 2629

\bibitem[Evrard et al.(2008)]{2008ApJ...672..122E} 
Evrard, A.~E., Bialek, J., Busha, M., et al.\ 2008, \apj, 672, 122 

\bibitem[Garmire et al.(2003)]{2003SPIE.4851...28G} 
Garmire, G.~P., Bautz, M.~W., Ford, P.~G., Nousek, J.~A., \& Ricker, G.~R., Jr.\ 2003, \procspie, 4851, 28 

\bibitem[\protect\citeauthoryear{Gehrels et al.}{2004}]{2004ApJ...611.1005G} 
Gehrels N., et al., 2004, ApJ, 611, 1005 

\bibitem[Gifford \& Miller(2013)]{2013ApJ...768L..32G} 
Gifford, D., \& Miller, C.~J.\ 2013, \apjl, 768, L32 

\bibitem[Gifford et al.(2013)]{2013ApJ...773..116G} 
Gifford, D., Miller, C., \& Kern, N.\ 2013, \apj, 773, 116 


\bibitem[Hicks et al.(2013)]{2013MNRAS.431.2542H} 
Hicks, A.~K., Pratt, G.~W., Donahue, M., et al.\ 2013, \mnras, 431, 2542 

\bibitem[Hill \& Oegerle(1993)]{1993AJ....106..831H} 
Hill, J.~M., \& Oegerle, W.~R.\ 1993, \aj, 106, 831 

\bibitem[Jones et al.(2009)]{2009MNRAS.399..683J} 
Jones, D.~H., Read, M.~A., Saunders, W., et al.\ 2009, \mnras, 399, 683 

\bibitem[Kravtsov et al.(2006)]{2006ApJ...650..128K} 
Kravtsov, A.~V., Vikhlinin, A., \& Nagai, D.\ 2006, \apj, 650, 128 

\bibitem[Liske et al.(2015)]{2015MNRAS.452.2087L} 
Liske, J., Baldry, I.~K., Driver, S.~P., et al.\ 2015, \mnras, 452, 2087 

\bibitem[Maughan(2007)]{2007ApJ...668..772M} 
Maughan, B.~J.\ 2007, \apj, 668, 772 

\bibitem[Maughan et al.(2012)]{2012MNRAS.421.1583M} 
Maughan, B.~J., Giles, P.~A., Randall, S.~W., Jones, C., \& Forman, W.~R.\ 2012, \mnras, 421, 1583 

\bibitem[Miller et al.(2002)]{2002AJ....124.1918M} 
Miller, C.~J., Krughoff, K.~S., Batuski, D.~J., \& Hill, J.~M.\ 2002, \aj, 124, 1918 

\bibitem[Miller et al.(2005)]{2005AJ....130..968M} 
Miller, C.~J., Nichol, R.~C., Reichart, D., et al.\ 2005, \aj, 130, 968 

\bibitem[Moretti et al.(2009)]{2009A&A...493..501M} 
Moretti, A., Pagani, C., Cusumano, G., et al.\ 2009, \aap, 493, 501 


\bibitem[Munari et al.(2013)]{2013MNRAS.430.2638M} 
Munari, E., Biviano, A., Borgani, S., Murante, G., \& Fabjan, D.\ 2013, \mnras, 430, 2638 

\bibitem[Navarro et al.(1997)]{1997ApJ...490..493N} 
Navarro, J.~F., Frenk, C.~S., \& White, S.~D.~M.\ 1997, \apj, 490, 493 

\bibitem[Nagai(2006)]{2006ApJ...650..538N} 
Nagai, D.\ 2006, \apj, 650, 538 

\bibitem[Pacaud et al.(2007)]{2007MNRAS.382.1289P} 
Pacaud, F., Pierre, M., Adami, C., et al.\ 2007, \mnras, 382, 1289 

\bibitem[Planck Collaboration et al.(2011)]{2011A&A...536A...9P} 
Planck Collaboration, Aghanim, N., Arnaud, M., et al.\ 2011, \aap, 536, AA9 

\bibitem[Planck Collaboration et al.(2012)]{2012A&A...543A.102P} 
Planck Collaboration, Aghanim, N., Arnaud, M., et al.\ 2012, \aap, 543, AA102 

\bibitem[Planck Collaboration et al.(2015)]{2015arXiv150201598P} 
Planck Collaboration, Ade, P.~A.~R., Aghanim, N., et al.\ 2015, arXiv:1502.01598 

\bibitem[Pike et al.(2014)]{2014MNRAS.445.1774P} 
Pike, S.~R., Kay, S.~T., Newton, R.~D.~A., Thomas, P.~A., \& Jenkins, A.\ 2014, \mnras, 445, 1774 

\bibitem[Pimbblet et al.(2006)]{2006MNRAS.366..645P} 
Pimbblet, K.~A., Smail, I., Edge, A.~C., et al.\ 2006, \mnras, 366, 645 

\bibitem[Pratt et al.(2009)]{2009A&A...498..361P} 
Pratt, G.~W., Croston, J.~H., Arnaud, M., Bohringer, H.\ 2009, \aap, 498, 361 

\bibitem[Rines et al.(2013)]{2013ApJ...767...15R} 
Rines, K., Geller, M.~J., Diaferio, A., \& Kurtz, M.~J.\ 2013, \apj, 767, 15 

\bibitem[Rosati et al.(2002)]{2002ARA&A..40..539R} 
Rosati, P., Borgani, S., \& Norman, C.\ 2002, \araa, 40, 539 

\bibitem[Rowley et al.(2004)]{2004MNRAS.352..508R} 
Rowley, D.~R., Thomas, P.~A., \& Kay, S.~T.\ 2004, \mnras, 352, 508 


\bibitem[Serra et al.(2011)]{2011MNRAS.412..800S} 
Serra, A.~L., Diaferio, A., Murante, G., \& Borgani, S.\ 2011, \mnras, 412, 800 

\bibitem[Stanek et al.(2006)]{2006ApJ...648..956S} 
Stanek, R., Evrard, A.~E., B{\"o}hringer, H., Schuecker, P., \& Nord, B.\ 2006, \apj, 648, 956 

\bibitem[Svensmark et al.(2015)]{2015MNRAS.448.1644S} 
Svensmark, J., Wojtak, R., \& Hansen, S.~H.\ 2015, \mnras, 448, 1644 

\bibitem[Turner et al.(2001)]{2001A&A...365L..27T} 
Turner, M.~J.~L., Abbey, A., Arnaud, M., et al.\ 2001, \aap, 365, L27 

\bibitem[Vikhlinin et al.(2009)]{2009ApJ...692.1033V} 
Vikhlinin, A., Burenin, R.~A., Ebeling, H., et al.\ 2009, \apj, 692, 1033 

\bibitem[Vikhlinin et al.(2006)]{2006ApJ...640..691V} 
Vikhlinin, A., Kravtsov, A., Forman, W., et al.\ 2006, \apj, 640, 691 

\bibitem[Voit(2005)]{2005RvMP...77..207V} 
Voit, G.~M.\ 2005, Reviews of Modern Physics, 77, 207 

\bibitem[Voit \& Donahue(2005)]{2005ApJ...634..955V} 
Voit, G.~M., \& Donahue, M.\ 2005, \apj, 634, 955 

\bibitem[Wojtak et al.(2011)]{2011Natur.477..567W} 
Wojtak, R., Hansen, S.~H., \& Hjorth, J.\ 2011, \nat, 477, 567 

\bibitem[Young et al.(2011)]{2011MNRAS.413..691Y} 
Young, O.~E., Thomas, 
P.~A., Short, C.~J., \& Pearce, F.\ 2011, \mnras, 413, 691 
\end{thebibliography}
\end{document}